\documentclass[aps,pra,groupedaddress]{revtex4}
\usepackage{graphicx}
\begin{document}


\title{Generation of highly non-classical $n$-photon polarization
states\\ by super-bunching at a photon bottleneck}

\author{Holger F. Hofmann}
\affiliation{PRESTO,
Japan Science and Technology Corporation
(JST)\\
Research Institute for Electronic Science, Hokkaido 
University\\
Kita-12 Nishi-6, Kita-ku, Sapporo 060-0812, Japan}
\altaffiliation{present address:
Graduate School of Advanced Sciences of Matter, Hiroshima University,
Kagamiyama 1-3-1, Higashi Hiroshima 739-8530, Japan}
\email{h.hofmann@osa.org}
\date{\today}

\begin{abstract}
It is shown that coherent superpositions of two oppositely
polarized $n$-photon states can be created by post-selecting 
the transmission of $n$ independently generated photons
into a single mode transmission line. 
It is thus possible to generate highly non-classical 
$n$-photon polarization states using only the bunching
effects associated with the bosonic nature of photons.
The effects of mode-matching errors are discussed and 
the possibility of creating $n$-photon entanglement by
redistributing the photons into $n$ separate modes is
considered.
\end{abstract}


\maketitle

\section{introduction}

The creation of highly non-classical states is one of the
fundamental challenges in quantum optics. In particular,
multi-photon entanglement and superpositions of macroscopically
distinguishable states (commonly referred to as cat states, after 
Schr\"odinger's famous cat paradox \cite{Sch35}) 
may be very useful as resources for optical quantum 
information processes such as
teleportation, cloning, or quantum computation 
\cite{Mur00,Loo00,Mur99,Got99,Ral03,Bar03}.
Recently, it has been shown that multi-photon entanglement
can indeed be created and manipulated using only single photon
sources, beam splitters, and post-selection based on precise
photon detection \cite{Kni01,Ral01,Hof02a,Ral02,Pit03,Hof02b,
Zou02,Fiu03}. In previous investigations,
these methods have been applied to photonic qubits, where the
goal was to obtain exactly one photon per spatial mode.
In order to achieve this kind of output , it is necessary to 
discard the cases where several photons bunch up in a single 
spatial mode as unwanted errors. However, it is also possible 
to specifically select cases where several photons bunch up 
in the same mode. In particular, this method has been used 
to propose the generation of spatial mode entanglement 
\cite{Kok02,Fiu02}.

In the following, a related proposal is presented for the 
generation of highly non-classical polarization states.
It is shown that a coherent superposition of two 
oppositely polarized $n$-photon polarization states can be
obtained by transmitting $n$ independently generated photons 
with homogeneously distributed polarizations into a single
spatial mode. For large photon numbers, this $n$-photon 
polarization state has the non-classical statistical properties
of a cat state, since the superposition is between two well 
separated regions of the Poincare sphere \cite{Lui02}
and the two polarization states can be distinguished
by measuring only a few photons. On the other hand, highly
non-classical interference effects between the two components
of the superposition will be observable in the polarization
statistics of the Stokes vector components orthogonal to the 
polarization along which the superposition is prepared
\cite{Hof03}.

Once a cat-state superposition of polarization states is
realized in single mode, it is also possible to generate
the corresponding multi-particle entangled state by
redistributing the photons into separate channels, effectively
transforming the local state of $n$ photons in one spatial
mode into an entangled state of $n$ photons in $n$ spatial
modes. In the present proposal, the photons are then 
transferred from $n$ input modes to $n$ output 
modes through a single mode bottleneck. The quantum 
interference effects associated with the bosonic nature of 
photons leads to a super-bunching effect in the polarization,
resulting in maximal $n$-photon entanglement in the output.
It is thus possible to realize a strong interaction between
an arbitrarily large number of photons by temporarily bunching 
them into a single mode channel. The super-bunching effect
at a photon bottleneck may therefore be a useful tool in the 
realization of a wide range of multi-photon quantum operations.

\section{The super-bunching effect}

The bunching effect used to obtain the non-classical polarization
state can be understood by considering the analogy
with two photon bunching. If a horizontally polarized photon
and a vertically polarized photon are transmitted into the same
spatial mode, their circular polarizations will always be the
same - either both right polarized, or both left polarized. 
Quantum interference removes the component with different 
circular polarizations. This effect can be generalized to
$n$ photons by choosing a homogeneous distribution of linear 
polarization angles.
The quantum interference between the different linear single
photon polarizations then removes all components of the
circular polarization, except for the two components where
all photons have the same circular polarization.
However, quantum coherence
between the two maximally polarized cases is preserved 
since the bunching effect does not distinguish between 
right or left circular polarization.

The validity of this argument can be verified by 
defining the following $n$-photon input state,
\begin{equation}
\label{eq:in}
\mid \psi_{\mbox{in}} \rangle
= 2^{-(n/2)} \prod_{l=0}^{n-1} \left(\hat{a}^\dagger_R(l) 
+ \exp[-i\frac{2\pi}{n} l]\;\hat{a}^\dagger_L(l)\right)
\mid \mbox{vac.} \rangle,
\end{equation}
where $\hat{a}^\dagger_R(l)$ and $\hat{a}^\dagger_L(l)$
are the creation operators of right and left circular
polarization in the spatial mode $l$. The rotation of
the linear polarization in the input ports $l$ is thus
represented by the phases of $2 \pi l/n$ in the superposition
of the circular polarization components.
Linear optics can then be used to transfer all $n$ input modes
in a single mode output port described by
\begin{eqnarray}
\hat{b}_R &=& \frac{1}{\sqrt{n}}\sum_{l=0}^{n-1} \hat{a}_R(l)
\nonumber \\
\hat{b}_L &=& \frac{1}{\sqrt{n}}\sum_{l=0}^{n-1} \hat{a}_L(l).
\end{eqnarray}
Each input mode $\hat{a}_{R/L}(l)$ thus contributes equally
to the correspondingly polarized output mode $\hat{b}_{R/L}$.
Photon losses to output modes other than $\hat{b}_{R/L}$ can
be avoided by post-selection using high quantum efficiency
photon detectors in all unused output ports. If any photon
losses are detected, the output is discarded. This post-selection
procedure can then be represented by a projection operator
$\hat{P}$. The non-normalized post-selected output state then reads
\begin{eqnarray}
\label{eq:post}
\mid \psi_{\mbox{out}} \rangle &=& \hat{P} 
\mid \psi_{\mbox{in}} \rangle 
\nonumber \\
&=& (2 n)^{-n/2} \prod_{l=0}^{n-1} \left(\hat{b}^\dagger_R
+ \exp[-i\frac{2\pi}{n} l]\;\hat{b}^\dagger_L\right)
\mid \mbox{vac.} \rangle.
\end{eqnarray}
Since the post-selection condition represented by $\hat{P}$
has made the photons from different input ports indistinguishable
in the output, the terms with the same overall number of
right and left circular polarized photons now interfere with
each other. Because all sums over varying phase factors 
$\exp[-i 2\pi l/n]$ are zero, the only two remaining terms in
the output are the component with only right circular polarized
photons and the component with only left circular polarized photons,
\begin{eqnarray}
\label{eq:output}
\mid \psi_{\mbox{out}} \rangle 
&=& (2 n)^{-n/2} \left(\left( \hat{b}^\dagger_R\right)^n
+ \exp[-i\pi (n-1)] \left(\hat{b}^\dagger_L\right)^n\right)
\mid \mbox{vac.} \rangle
\nonumber \\
&=& \sqrt{\frac{n!}{(2n)^n}}\left(\mid n; 0 \rangle - (-1)^n
\mid 0; n \rangle\right),
\end{eqnarray}
where the final result is expressed in the photon number basis of
the right and left circular polarization modes in the output.
Note that the elimination of all other components of circular
polarization can also be justified by a symmetry argument.
Since the distribution of input polarizations has an $n$-fold
symmetry with respect to rotations of the Stokes vector around
the circular polarization axis (or a $(2n)$-fold rotation
symmetry of the linear polarization), and since the operations
applied are not sensitive to the polarization at all, the
output state must also have this $n$-fold symmetry. 
However, any coherent superposition of the maximally polarized 
states $\mid n; 0 \rangle$ and $\mid 0; n \rangle$ with other 
polarization states would reduce this symmetry \cite{Hof03}.
Therefore, the cat-state superposition of $\mid n; 0 \rangle$ and 
$\mid 0; n \rangle$ is the only possible output state that 
preserves the $n$-fold symmetry of the input. 

In post-selection methods, the difficulty of generating 
highly non-classical states is directly reflected in the
probability of obtaining the post-selection condition
that indicates a successful generation of the target state.
In the present case, this condition is given by the 
probability that no photons are detected in any spatial 
modes other than the output modes $\hat{b}_{R/L}$. 
The probability of success for this post-selection procedure
is given by 
\begin{equation}
p(n)= \langle \psi_{\mbox{out}} \mid \psi_{\mbox{out}} \rangle
= \frac{2 \, n!}{(2 n)^n}.
\end{equation}
Since the probability of finding all photons in the same 
spatial output mode rapidly decreases with photon number, the
efficiency of generating highly non-classical superpositions
is very low. However, the probability of success is still
significantly higher than the probability of finding all
of $n$ independent particles in the output channel. For example,
the probability of finding three photons in the output
is $p(3)=1/18$. For three independent particles, the chance
of finding all three in the output channel would be only
$(1/3)^3 = 1/27$. This enhancement of the post-selection
probability increases with increasing photon number and
may thus be helpful in the suppression of errors caused by
imperfect mode matching between the input photons (see below).

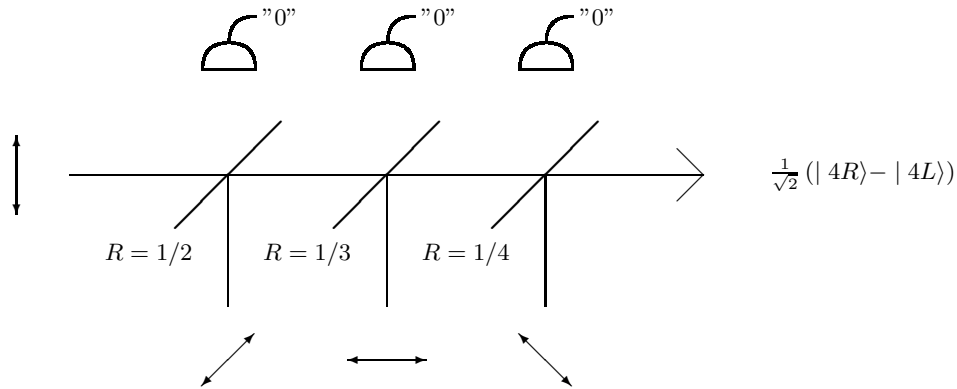
\begin{figure}
\vspace{-1cm}
\begin{picture}(480,240)



\put(80,120){\line(1,0){240}}
\put(320,120){\line(-1,1){10}}
\put(320,120){\line(-1,-1){10}}
\put(140,70){\line(0,1){50}}
\put(200,70){\line(0,1){50}}
\put(260,70){\line(0,1){50}}


\put(60,120){\vector(0,1){15}}
\put(60,120){\vector(0,-1){15}}

\put(140,50){\vector(1,1){10}}
\put(140,50){\vector(-1,-1){10}}

\put(200,50){\vector(1,0){15}}
\put(200,50){\vector(-1,0){15}}

\put(260,50){\vector(-1,1){10}}
\put(260,50){\vector(1,-1){10}}

\thicklines


\put(120,100){\line(1,1){40}}
\put(80,80){\makebox(60,20){$R=1/2$}}
\put(180,100){\line(1,1){40}}
\put(140,80){\makebox(60,20){$R=1/3$}}
\put(240,100){\line(1,1){40}}
\put(200,80){\makebox(60,20){$R=1/4$}}


\put(130,160){\line(1,0){20}}
\bezier{80}(130,160)(130,170)(140,170)
\bezier{80}(140,170)(150,170)(150,160)
\bezier{80}(140,170)(140,180)(150,180)
\put(150,170){\makebox(20,20){"0"}}

\put(190,160){\line(1,0){20}}
\bezier{80}(190,160)(190,170)(200,170)
\bezier{80}(200,170)(210,170)(210,160)
\bezier{80}(200,170)(200,180)(210,180)
\put(210,170){\makebox(20,20){"0"}}

\put(250,160){\line(1,0){20}}
\bezier{80}(250,160)(250,170)(260,170)
\bezier{80}(260,170)(270,170)(270,160)
\bezier{80}(260,170)(260,180)(270,180)
\put(270,170){\makebox(20,20){"0"}}

\put(320,110){\makebox(120,20){$\frac{1}{\sqrt{2}}
\left(\mid 4R \rangle - \mid 4L \rangle\right)$}}

\end{picture}
\vspace{-0.7cm}
\caption{\label{fig1} Schematic setup for the generation
of a four photon polarization cat state. Each input port
receives one photon with the linear polarization indicated
above. The detectors ensure that no photons are emitted 
into the empty output ports, and the beam splitter 
reflectivities are chosen so that the each input component 
has equal weight in the output.}
\end{figure}
Figure \ref{fig1} shows an example of an optical setup
to generate a cat state superposition of four photon 
polarization. The cascade setup shown can easily be
generalized to arbitrary photon numbers. Each input photon
is generated by a single photon source with a well defined
linear polarization, as indicated in the figure. The
post-selection condition is that no photons are detected in
the three detectors set up at the empty output ports.
Note that it would also be sufficient to post-select the
arrival of all four photons at detectors in the output. This 
may be useful to avoid errors due to the limited quantum
efficiencies of the detectors, although it might restrict 
the possibilities of further quantum operations on the output.

\section{Polarization statistics of the output state}

The polarization statistics of the output
can be characterized by the Stokes parameters, 
defined as the photon number difference between a pair of
orthogonal polarizations. The properties of the
quantized Stokes parameters then correspond to the 
properties of the spin components of a spin-$n/2$ system 
\cite{Hof03}. The quantum statistics of four photon 
polarization thus corresponds to a spin-2 system, and
the five possible photon distributions between any two
orthogonal polarizations correspond to the five eigenvalues
of the respective spin component.

The most impressive feature of the super-bunching
effect is the accumulation of all photons in the two states
with maximal circular polarization, given by the circular
polarization statistics $p_{RL}(n_R,n_L)$,
\begin{eqnarray}
\label{eq:bunchstat}
p_{RL}(4;0) &=& 1/2
\nonumber \\
p_{RL}(3;1) &=& 0
\nonumber \\
p_{RL}(2;2) &=& 0
\nonumber \\
p_{RL}(1;3) &=& 0
\nonumber \\
p_{RL}(0;4) &=& 1/2.
\end{eqnarray}
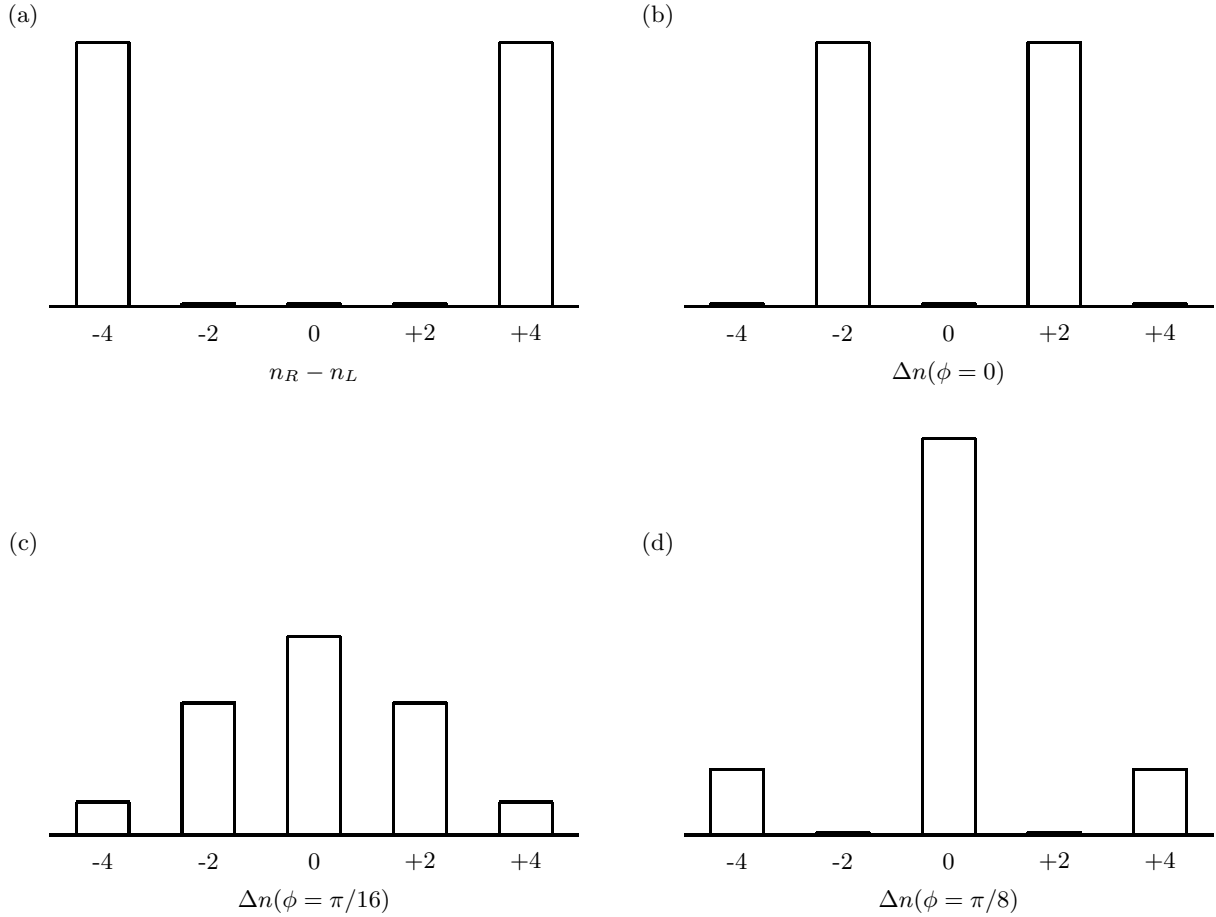
\begin{figure}
\begin{picture}(480,400)
\thicklines


\put(0,340){\makebox(20,20){(a)}}

\put(20,240){\line(1,0){200}}

\put(30,240){\line(0,1){100}}
\put(30,340){\line(1,0){20}}
\put(50,240){\line(0,1){100}}
\put(30,220){\makebox(20,20){-4}}

\put(70,240){\line(0,1){1}}
\put(70,241){\line(1,0){20}}
\put(90,240){\line(0,1){1}}
\put(70,220){\makebox(20,20){-2}}

\put(110,240){\line(0,1){1}}
\put(110,241){\line(1,0){20}}
\put(130,240){\line(0,1){1}}
\put(110,220){\makebox(20,20){0}}

\put(150,240){\line(0,1){1}}
\put(150,241){\line(1,0){20}}
\put(170,240){\line(0,1){1}}
\put(150,220){\makebox(20,20){+2}}

\put(190,240){\line(0,1){100}}
\put(190,340){\line(1,0){20}}
\put(210,240){\line(0,1){100}}
\put(190,220){\makebox(20,20){+4}}

\put(80,205){\makebox(80,20){$n_R-n_L$}}


\put(240,340){\makebox(20,20){(b)}}

\put(260,240){\line(1,0){200}}

\put(270,240){\line(0,1){1}}
\put(270,241){\line(1,0){20}}
\put(290,240){\line(0,1){1}}
\put(270,220){\makebox(20,20){-4}}

\put(310,240){\line(0,1){100}}
\put(310,340){\line(1,0){20}}
\put(330,240){\line(0,1){100}}
\put(310,220){\makebox(20,20){-2}}

\put(350,240){\line(0,1){1}}
\put(350,241){\line(1,0){20}}
\put(370,240){\line(0,1){1}}
\put(350,220){\makebox(20,20){0}}

\put(390,240){\line(0,1){100}}
\put(390,340){\line(1,0){20}}
\put(410,240){\line(0,1){100}}
\put(390,220){\makebox(20,20){+2}}

\put(430,240){\line(0,1){1}}
\put(430,241){\line(1,0){20}}
\put(450,240){\line(0,1){1}}
\put(430,220){\makebox(20,20){+4}}

\put(320,205){\makebox(80,20){$\Delta n (\phi=0)$}}


\put(0,140){\makebox(20,20){(c)}}

\put(20,40){\line(1,0){200}}

\put(30,40){\line(0,1){12.5}}
\put(30,52.5){\line(1,0){20}}
\put(50,40){\line(0,1){12.5}}
\put(30,20){\makebox(20,20){-4}}

\put(70,40){\line(0,1){50}}
\put(70,90){\line(1,0){20}}
\put(90,40){\line(0,1){50}}
\put(70,20){\makebox(20,20){-2}}

\put(110,40){\line(0,1){75}}
\put(110,115){\line(1,0){20}}
\put(130,40){\line(0,1){75}}
\put(110,20){\makebox(20,20){0}}

\put(150,40){\line(0,1){50}}
\put(150,90){\line(1,0){20}}
\put(170,40){\line(0,1){50}}
\put(150,20){\makebox(20,20){+2}}

\put(190,40){\line(0,1){12.5}}
\put(190,52.5){\line(1,0){20}}
\put(210,40){\line(0,1){12.5}}
\put(190,20){\makebox(20,20){+4}}

\put(80,5){\makebox(80,20){$\Delta n (\phi=\pi/16)$}}


\put(240,140){\makebox(20,20){(d)}}

\put(260,40){\line(1,0){200}}

\put(270,40){\line(0,1){25}}
\put(270,65){\line(1,0){20}}
\put(290,40){\line(0,1){25}}
\put(270,20){\makebox(20,20){-4}}

\put(310,40){\line(0,1){1}}
\put(310,41){\line(1,0){20}}
\put(330,40){\line(0,1){1}}
\put(310,20){\makebox(20,20){-2}}

\put(350,40){\line(0,1){150}}
\put(350,190){\line(1,0){20}}
\put(370,40){\line(0,1){150}}
\put(350,20){\makebox(20,20){0}}

\put(390,40){\line(0,1){1}}
\put(390,41){\line(1,0){20}}
\put(410,40){\line(0,1){1}}
\put(390,20){\makebox(20,20){+2}}

\put(430,40){\line(0,1){25}}
\put(430,65){\line(1,0){20}}
\put(450,40){\line(0,1){25}}
\put(430,20){\makebox(20,20){+4}}

\put(320,5){\makebox(80,20){$\Delta n (\phi=\pi/8)$}}

\end{picture}

\caption{\label{catstat} Illustration of the characteristic polarization statistics of
a superposition state of the two four photon states with maximal circular polarization,
$\mid 4 R \rangle$ and $\mid 4 L \rangle$. (a) shows the circular polarization statistics,
(b) shows the linear polarization statistics along the horizontal and vertical directions
($\phi=0$), (c) shows the polarization statistics at an angle of $\pi/16$ relative to
the horizontal and vertical directions, and (d) shows the polarization statistics at an
angle of $\pi/8$. Note that the linear polarization statistics have a periodicity of
$\pi/4$, that is, the statistics shown in (a) also applies to an angle of $\pi/4$, etc.
}

\end{figure}
However, the observation of this bunching effect around
$n_R-n_L=\pm 4$ does not indicate coherence between the
two components. To distinguish the quantum superposition
from a statistical mixture, it is necessary to consider
the linear polarization statistics. These can be obtained
from the coherent overlap of the two components
$\mid 4R \rangle$ and $\mid 4L \rangle$ with the basis
states of a linear polarization measurement rotated
by an angle of $\phi$ relative to the horizontal and
vertical polarization axes, where the individual basis
states are defined by the photon number difference
$\Delta n (\phi)$ between the two orthogonal
polarizations,
\begin{equation}
\begin{array}{lcrclcr}
\langle  \Delta n = +4 \mid 4R \rangle
    &=& 1/4 \exp[-i 4 \phi],
&\hspace{1cm}& \langle  \Delta n = +4 \mid 4L \rangle
    &=& 1/4 \exp[+i 4 \phi],
\\
\langle  \Delta n = +2 \mid 4R \rangle
    &=& 1/2 \exp[-i 4 \phi],
&& \langle  \Delta n = +2 \mid 4L \rangle
    &=& - 1/2 \exp[+i 4 \phi],
\\
\langle  \Delta n = \; \; 0 \; \mid 4R \rangle
    &=& \sqrt{6}/4 \exp[-i 4 \phi],
&& \langle  \Delta n = \; \; 0 \; \mid 4L \rangle
    &=& \sqrt{6}/4 \exp[+i 4 \phi],
\\
\langle  \Delta n = -2 \mid 4R \rangle
    &=& 1/2 \exp[-i 4 \phi],
&& \langle  \Delta n = -2 \mid 4L \rangle
    &=& - 1/2 \exp[+i 4 \phi],
\\
\langle  \Delta n = -4 \mid 4R \rangle
    &=& 1/4 \exp[-i 4 \phi],
&& \langle  \Delta n = -4 \mid 4L \rangle
    &=& 1/4 \exp[+i 4 \phi].
\end{array}
\end{equation}
As the linear polarization is rotated, the interference terms in the photon number
distribution thus oscillate with a periodicity of $8 \phi$,
\begin{eqnarray}
\label{eq:linphi}
p_\phi (+4) &=& \frac{1}{16}\left(1-\cos[8 \phi]\right),
\nonumber \\
p_\phi (+2) &=& \frac{4}{16}\left(1+\cos[8 \phi]\right),
\nonumber \\
p_\phi (0) &=& \frac{6}{16}\left(1-\cos[8 \phi]\right),
\nonumber \\
p_\phi (-2) &=& \frac{4}{16}\left(1+\cos[8 \phi]\right),
\nonumber \\
p_\phi (-4) &=& \frac{1}{16}\left(1-\cos[8 \phi]\right).
\end{eqnarray}
Figure \ref{catstat} shows the characteristic polarization statistics of the super-bunched
superposition state. Fig. \ref{catstat} (a) shows the super-bunching effect in the circular
polarization statistics, fig. \ref{catstat} (b) to (d) show the linear polarization statistics
for polarization angles of $\phi=0$, $\phi=\pi/8$, and $\phi=\pi/16$, respectively.
Note that the polarization angles are defined relative to the four input polarizations,
that is, the sensitivity of the output statistics to the linear polarization direction
originates from the anisotropy caused by the selection of a specific set of
input polarizations. Specifically, the polarization statistics of linear polarizations
that coincide with one of the input polarizations ($\phi=0$ and $\phi=\pi/4$)
always have exactly three photons in one polarization and one in the other ($\Delta n (\phi) = \pm 2 $), as can be seen in fig. \ref{catstat} (b). This distribution of
photons can be understood as a combination of the preservation of the input polarization (one input photon in each of the two observed output polarization) and the
conventional bunching effect
between the remaining two photons, which adds exactly two photons to either
one of the two output polarizations. On the other hand, the polarization
statistics of linear polarizations that are exactly halfway between two
of the input polarizations ($\phi=\pi/8$ and $\phi=3 \pi/8$) are clearly dominated
by the 75\% probability of finding equal numbers of photons in both
polarizations ($\Delta n (\phi) = 0 $), as can be seen in fig. \ref{catstat} (d).
In this case, the
complete absence of
photon distributions with $\Delta n (\phi) = \pm 2 $ indicates a strong quantum
interference effect between the circular polarization components. The linear
polarization statistics that would also be expected for the maximally right or
left polarized states $\mid 4 R \rangle$ or $\mid 4 L \rangle$ are obtained
at angles of $\phi=\pi/16$, $\phi=3\pi/16$, etc., as shown in fig. \ref{catstat} (c). These statistics correspond to the binomial distribution expected if each photon had a random linear polarization. The deviations from this binomial
distribution thus show that the polarization statistics cannot be explained in terms of a simple combination of individual photon polarizations.

\section{Effects of mode matching errors}

Experimentally, the super-bunching effect requires that all input photons can be matched
into the same mode at the beam splitters. Therefore, the most likely source of errors
is an imperfect mode matching at one of the input ports. Since the effect of such errors
is quite different from the decoherence effects normally expected in spin systems, it
may be of interest to investigate the polarization statistics associated with a mode
matching error in more detail. 

The characteristics of a mode matching error can be obtained by assuming that one of
the photons is in a different mode from the other three. The output polarization state
is then a product state of a three photon state with modified bunching effects and
a single photon state with the unchanged input polarization.
If the mismatched photon is horizontally polarized (represented by the
contribution with $l=0$ in equation (\ref{eq:in})),
the output state of the remaining three photons is given by
\begin{eqnarray}
\label{eq:partbunch}
\frac{1}{16 \sqrt{2}}(\hat{a}^\dagger_R+i \hat{a}^\dagger_L)
(\hat{a}^\dagger_R- \hat{a}^\dagger_L)
(\hat{a}^\dagger_R- i\hat{a}^\dagger_L) \mid \mbox{vac.}\rangle
&=&
  \frac{\sqrt{3}}{16} \mid 3;0 \rangle 
- \frac{1}{16} \mid 2;1 \rangle
+ \frac{1}{16} \mid 1;2 \rangle
- \frac{\sqrt{3}}{16} \mid 0;3 \rangle. 
\end{eqnarray}
Note that the circular polarization statistics of this state is independent
of the linear polarization of the mismatched photon. Therefore, all mode matching
errors modify the circular polarization in the same way. However, the loss of symmetry
will be apparent in the linear polarization statistics.

The total output state is obtained as a product state of the bunched three photon
state given by equation (\ref{eq:partbunch}) and the single horizontally polarized photon
transmitted to the output port with a probability of $1/4$. 
In the circular polarization basis, this product state of three photon and one photon
polarization in the output reads
\begin{eqnarray}
\label{eq:mmRL}
\mid \psi_{3 \otimes 1}\rangle &=& \frac{1}{32 \sqrt{2}}
\left( \sqrt{3} \mid 3;0 \rangle 
- \mid 2;1 \rangle
+ \mid 1;2 \rangle
- \sqrt{3} \mid 0;3 \rangle \right) \otimes \left(\mid 1;0 \rangle + \mid 0;1 \rangle \right). 
\end{eqnarray}
The total amplitude of this state describes the post-selection probability. It is interesting
to compare this probability to the perfectly mode matched four photon post selection 
probability $p(4)$, 
\begin{equation}
\label{eq:mmnorm}
\langle \psi_{3 \otimes 1} \mid \psi_{3 \otimes 1} \rangle
= \frac{1}{128} = \frac{2}{3} \; p(4).
\end{equation}
The mode mismatch thus reduces the post-selection probability for the four photons
from $3/256$ to $2/256$. Experimentally, this dependence
of successful post-selection on mode matching could be observed 
by varying the delay time of one of the input photons, thus 
varying the phase matching artificially \cite{San04}. It is then possible to 
evaluate the mode matching from the rates of coincidence counts
obtained in the experiment.

The statistics of circular polarization obtained from a single photon mismatch
error corresponding to the output state $\mid \psi_{3 \otimes 1} \rangle$ reads
\begin{eqnarray}
\label{eq:error}
p_{3 \otimes 1}(4;0) &=& \frac{3}{16}
\nonumber \\
p_{3 \otimes 1}(3;1) &=& \frac{4}{16}
\nonumber \\
p_{3 \otimes 1}(2;2) &=& \frac{2}{16}
\nonumber \\
p_{3 \otimes 1}(1;3) &=& \frac{4}{16}
\nonumber \\
p_{3 \otimes 1}(0;4) &=& \frac{3}{16}.
\end{eqnarray}
This distribution sharply contrasts the bunching effect of the perfectly 
mode matched case given by equation (\ref{eq:bunchstat}). The main effect of the
mode matching error is thus to destroy the bunching effect, while the coherence 
between the circularly polarized components, which corresponds to the linear
polarization statistics, is affected much less.

\begin{figure}
\begin{picture}(480,200)
\thicklines


\put(0,140){\makebox(20,20){(a)}}

\put(20,40){\line(1,0){200}}

\put(30,40){\line(0,1){30}}
\put(30,70){\line(1,0){20}}
\put(50,40){\line(0,1){30}}
\put(30,20){\makebox(20,20){-4}}

\put(70,40){\line(0,1){40}}
\put(70,80){\line(1,0){20}}
\put(90,40){\line(0,1){40}}
\put(70,20){\makebox(20,20){-2}}

\put(110,40){\line(0,1){20}}
\put(110,60){\line(1,0){20}}
\put(130,40){\line(0,1){20}}
\put(110,20){\makebox(20,20){0}}

\put(150,40){\line(0,1){40}}
\put(150,80){\line(1,0){20}}
\put(170,40){\line(0,1){40}}
\put(150,20){\makebox(20,20){+2}}

\put(190,40){\line(0,1){30}}
\put(190,70){\line(1,0){20}}
\put(210,40){\line(0,1){30}}
\put(190,20){\makebox(20,20){+4}}

\put(80,5){\makebox(80,20){$n_R-n_L$}}


\put(240,140){\makebox(20,20){(b)}}

\put(260,40){\line(1,0){200}}

\put(270,40){\line(0,1){1}}
\put(270,41){\line(1,0){20}}
\put(290,40){\line(0,1){1}}
\put(270,20){\makebox(20,20){-4}}

\put(310,40){\line(0,1){120}}
\put(310,160){\line(1,0){20}}
\put(330,40){\line(0,1){120}}
\put(310,20){\makebox(20,20){-2}}

\put(350,40){\line(0,1){1}}
\put(350,41){\line(1,0){20}}
\put(370,40){\line(0,1){1}}
\put(350,20){\makebox(20,20){0}}

\put(390,40){\line(0,1){40}}
\put(390,80){\line(1,0){20}}
\put(410,40){\line(0,1){40}}
\put(390,20){\makebox(20,20){+2}}

\put(430,40){\line(0,1){1}}
\put(430,41){\line(1,0){20}}
\put(450,40){\line(0,1){1}}
\put(430,20){\makebox(20,20){+4}}

\put(320,5){\makebox(80,20){$\Delta n (\phi=0)$}}

\thinlines


\put(35,40){\line(0,1){120}}
\put(35,160){\line(1,0){20}}
\put(55,40){\line(0,1){120}}

\put(75,40){\line(0,1){1}}
\put(75,41){\line(1,0){20}}
\put(95,40){\line(0,1){1}}

\put(115,40){\line(0,1){1}}
\put(115,41){\line(1,0){20}}
\put(135,40){\line(0,1){1}}

\put(155,40){\line(0,1){1}}
\put(155,41){\line(1,0){20}}
\put(175,40){\line(0,1){1}}

\put(195,40){\line(0,1){120}}
\put(195,160){\line(1,0){20}}
\put(215,40){\line(0,1){120}}


\put(275,40){\line(0,1){1}}
\put(275,41){\line(1,0){20}}
\put(295,40){\line(0,1){1}}

\put(315,40){\line(0,1){120}}
\put(315,160){\line(1,0){20}}
\put(335,40){\line(0,1){120}}

\put(355,40){\line(0,1){1}}
\put(355,41){\line(1,0){20}}
\put(375,40){\line(0,1){1}}

\put(395,40){\line(0,1){120}}
\put(395,160){\line(1,0){20}}
\put(415,40){\line(0,1){120}}

\put(435,40){\line(0,1){1}}
\put(435,41){\line(1,0){20}}
\put(455,40){\line(0,1){1}}

\end{picture}
\caption{\label{error} Illustration of the change in the 
polarization statistics caused by a one photon mode-matching error.
(a) shows the circular and (b) shows the horizontal/vertical
polarization statistics. The thick lines show the polarization statistics
of the error component, corresponding to three photons in the same mode 
and one photon in a different mode, as given by equations (\ref{eq:mmRL}) and  
(\ref{eq:mmHV}). 
The thin lines show the output probabilities for the ideal four-photon
bunching. The difference in the total probability corresponds to the 
change of post-selection probability}
\end{figure}
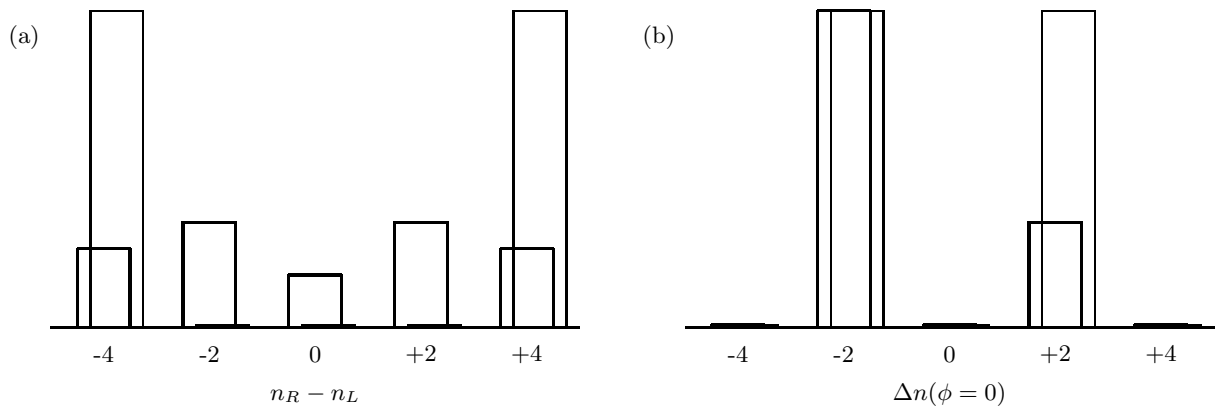

To evaluate the effects of the mode mismatch on the linear polarization
components, it is useful to transform the output state into the horizontal and
vertical basis states ($\phi=0$). In this basis, the state reads
\begin{eqnarray}
\label{eq:mmHV}
\mid \psi_{3 \otimes 1}\rangle &=& \frac{1}{16 \sqrt{2}}
\left( \sqrt{3} \mid 3V \rangle 
- \mid 2H;1V \rangle \right) \otimes \mid 1H \rangle. 
\end{eqnarray}
The horizontal/vertical polarization statistics thus still includes only the two
components with $\Delta n = \pm 2$, corresponding to three photons in one polarization
and one in the other. The only change compared to the statistics for $\phi=0$ 
given in equation (\ref{eq:linphi}) is that the vertically polarized output is preferred.

Figure \ref{error} shows a comparison of the polarization statistics for the one photon
mode mismatch with the perfectly mode matched four photon output. Note that the changed
post selection probability has been taken into account, so that the sum over all probabilities
shown is 1.5 times higher for the perfectly mode matched case. 
Most importantly, mode matching errors have more impact on the circular polarization
statistics than on the linear polarization statistics which depends on coherence between
the circular polarization eigenstates. The errors expected in the generation of a
superposition state of $\mid 4R \rangle$ and $\mid 4L \rangle$ by superbunching are
therefore quite different from the decoherence effects normally associated with cat-state
superpositions. In the case of a small mode matching error probability $\epsilon$,
the expected circular polarization distribution can be derived by mixing the ideal
distribution of equation (\ref{eq:bunchstat}) with the error distribution given in equation (\ref{eq:error}). Taking the different post-selection probabilities into account, this distribution
is given by
\begin{equation}
p_{\mbox{error}}(n;4-n) = \frac{3}{256} \; (1-\epsilon) p_{RL}(n;4-n)
+ \frac{1}{128} \; \epsilon \; p_{3\otimes1}(n;4-n).
\end{equation}
It is thus possible to estimate the mode matching error from the experimental data
obtained in measurements of the circular polarization output.

\section{Generation of multi-particle entanglement}

Once a highly non-classical $n$-photon polarization state is generated in
single mode, this state can be converted into an $n$-particle entangled state
by distributing the photons into $n$ separate channels. It is then possible
to apply the criteria for entanglement verification to quantify the
non-classical features of the polarization statistics.
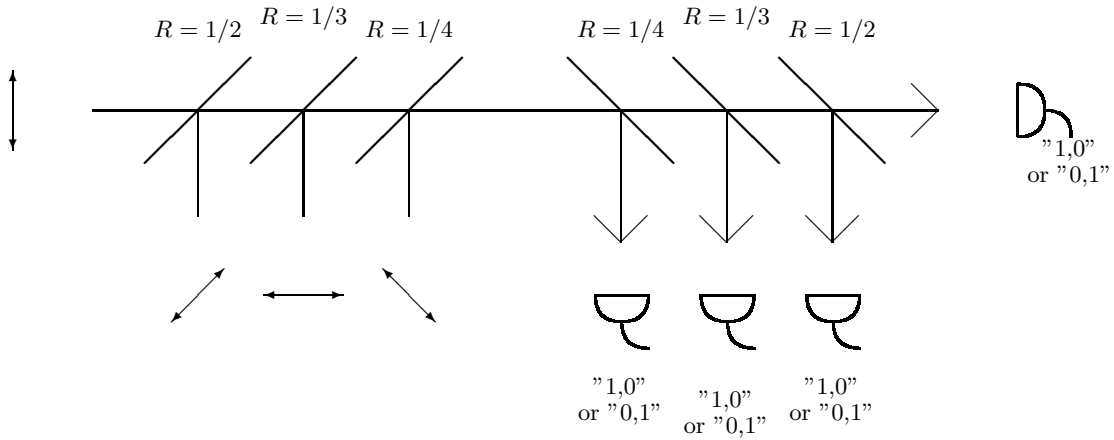
\begin{figure}
\begin{picture}(480,240)


\put(70,160){\line(1,0){320}}
\put(390,160){\line(-1,1){10}}
\put(390,160){\line(-1,-1){10}}
\put(110,120){\line(0,1){40}}
\put(150,120){\line(0,1){40}}
\put(190,120){\line(0,1){40}}
\put(270,110){\line(0,1){50}}
\put(270,110){\line(1,1){10}}
\put(270,110){\line(-1,1){10}}
\put(310,110){\line(0,1){50}}
\put(310,110){\line(1,1){10}}
\put(310,110){\line(-1,1){10}}
\put(350,110){\line(0,1){50}}
\put(350,110){\line(1,1){10}}
\put(350,110){\line(-1,1){10}}


\put(40,160){\vector(0,1){15}}
\put(40,160){\vector(0,-1){15}}

\put(110,90){\vector(1,1){10}}
\put(110,90){\vector(-1,-1){10}}

\put(150,90){\vector(1,0){15}}
\put(150,90){\vector(-1,0){15}}

\put(190,90){\vector(-1,1){10}}
\put(190,90){\vector(1,-1){10}}

\thicklines


\put(90,140){\line(1,1){40}}
\put(80,180){\makebox(60,20){$R=1/2$}}
\put(130,140){\line(1,1){40}}
\put(120,185){\makebox(60,20){$R=1/3$}}
\put(170,140){\line(1,1){40}}
\put(160,180){\makebox(60,20){$R=1/4$}}

\put(290,140){\line(-1,1){40}}
\put(240,180){\makebox(60,20){$R=1/4$}}
\put(330,140){\line(-1,1){40}}
\put(280,185){\makebox(60,20){$R=1/3$}}
\put(370,140){\line(-1,1){40}}
\put(320,180){\makebox(60,20){$R=1/2$}}


\put(260,90){\line(1,0){20}}
\bezier{80}(260,90)(260,80)(270,80)
\bezier{80}(270,80)(280,80)(280,90)
\bezier{80}(270,80)(270,70)(280,70)
\put(250,45){\makebox(40,20){"1,0"}}
\put(250,35){\makebox(40,20){or "0,1"}}

\put(300,90){\line(1,0){20}}
\bezier{80}(300,90)(300,80)(310,80)
\bezier{80}(310,80)(320,80)(320,90)
\bezier{80}(310,80)(310,70)(320,70)
\put(290,40){\makebox(40,20){"1,0"}}
\put(290,30){\makebox(40,20){or "0,1"}}

\put(340,90){\line(1,0){20}}
\bezier{80}(340,90)(340,80)(350,80)
\bezier{80}(350,80)(360,80)(360,90)
\bezier{80}(350,80)(350,70)(360,70)
\put(330,45){\makebox(40,20){"1,0"}}
\put(330,35){\makebox(40,20){or "0,1"}}

\put(420,150){\line(0,1){20}}
\bezier{80}(420,150)(430,150)(430,160)
\bezier{80}(430,160)(430,170)(420,170)
\bezier{80}(430,160)(440,160)(440,150)
\put(420,135){\makebox(40,20){"1,0"}}
\put(420,125){\makebox(40,20){or "0,1"}}

\end{picture}
\caption{\label{bottleneck} Schematic setup for the generation of
multi-particle entanglement using the photon bottleneck. In this
case, post-selection by polarization sensitive detection
in the four output channels is necessary to
redistribute the photons into different modes.}
\end{figure}
Figure \ref{bottleneck} illustrates a possible setup for the generation of
four photon entanglement using a photon bottleneck. In this setup, it is
necessary to post-select the output by detecting exactly one photon in
each output channel using polarization sensitive detectors.
The redistribution of the photons into
separate channels thus makes the output condition symmetric to the
input condition and the operation of the photon bottleneck can be
interpreted as a collective four photon interaction.
In the absence of errors,
the output state of this interaction is the GHZ-state,
\begin{equation}
\mid \mbox{GHZ} \rangle = \frac{1}{\sqrt{2}} \left(\mid RRRR \rangle -
\mid LLLL \rangle \right).
\end{equation}
Such four photon GHZ-states have recently been generated by using entangled
pairs generated in downconversion as input \cite{Pan01,Eib03,Zha03}.
The photon bottleneck provides an alternative method of generating the
same type of multi-particle entanglement from previously unentangled input photons.

The analysis of mode matching errors given above can now be applied to
determine the condition for successful entanglement generation.
Genuine $n$-particle entanglement can be verified by
the condition \cite{Guh03}
\begin{equation}
\frac{
\langle \mbox{GHZ} \mid
\hat{\rho}_{\mbox{out}}
\mid \mbox{GHZ} \rangle}
{\mbox{Tr}\left\{ \hat{\rho}_{\mbox{out}} \right\}}
\geq \frac{1}{2}.
\end{equation}
Using the results of equations (\ref{eq:mmRL}) and (\ref{eq:mmnorm}),
it is possible to quantify the reduction in multi-particle
entanglement caused by a mode matching error as follows,
\begin{equation}
\frac{
|\langle \mbox{GHZ} \mid \psi_{3 \otimes 1} \rangle|^2}
{\langle \psi_{3 \otimes 1} \mid \psi_{3 \otimes 1} \rangle}
= \frac{3}{8}.
\end{equation}
The overall reduction for small mode matching error probabilities
$\epsilon$ can then be determined using the different post-selection
probabilites as
\begin{equation}
\frac{
\langle \mbox{GHZ} \mid
\hat{\rho}_{\mbox{out}}
\mid \mbox{GHZ} \rangle}
{\mbox{Tr}\left\{ \hat{\rho}_{\mbox{out}} \right\}}
= \frac{12-9 \epsilon}{12-4 \epsilon} \approx 1-\frac{5}{12} \epsilon.
\end{equation}
For example, even if each of the four input channels contributes a
mode matching error of 5\%, for a total error probability of about
20\%  ($\epsilon \approx 0.2$), the GHZ contribution would only be
reduced by about 8.3 \%. The generation of multi-particle
entanglement using a photon bottleneck thus appears to be very
robust against typical mode matching errors.

\section{Conclusions}

In conclusion, it has been shown that a highly non-classical
superposition state of oppositely polarized $n$-photon 
states can be generated by post-selecting the transmission 
of $n$ independently generated photons into a single spatial 
mode. No initial entanglement is needed,
and the post-selection conditions only require zero detection
events. In principle, the method can be applied to any number
of photons. It can be used to generate cat-like 
superpositions in the polarization statistics of single
mode $n$-photon states, or to obtain GHZ-type $n$-photon 
entanglement. The error analysis suggests that the 
non-classical correlations that can be generated by this 
method are sufficiently robust with regard to experimental 
imperfections. The photon bottleneck setup presented in this
paper may therefore provide a useful tool for the generation 
and control of non-classical states of light.

\end{document}